\documentclass[aps,prl,superscriptaddress,reprint,twocolumn]{revtex4-2}

\usepackage{listings}
\usepackage{amsmath,amssymb, mathrsfs}
\usepackage{hyperref}
\hypersetup{pdfborder={0 0 0},colorlinks=true,urlcolor=blue,citecolor=blue,linkcolor=blue,linktocpage=true}

\begin{document}

\title{Holography for the IKKT matrix model}

\author{Franz Ciceri}
\affiliation{ENS de Lyon, CNRS, LPENSL, UMR5672, 69342, Lyon cedex 07, France}

\author{Henning Samtleben}
\affiliation{ENS de Lyon, CNRS, LPENSL, UMR5672, 69342, Lyon cedex 07, France}

\affiliation{Institut Universitaire de France (IUF)}

\begin{abstract}
\noindent Holographic dualities that relate type II strings on near-horizon D$p$ brane geometries to super Yang-Mills theories with sixteen supercharges in $p+1$ dimensions provide non-conformal generalizations of the famous AdS/CFT correspondence. For the extremal case $p=-1$, this suggests a holographic duality for the IKKT matrix model --- super Yang-Mills theory reduced to zero dimensions. Despite intriguing and highly non-trivial results in the IKKT model, this duality has remained largely unexplored so far. In this letter, we consider the lowest supermultiplet of gauge invariant operators of the model and identify its states with the lowest Kaluza-Klein fluctuations of (Euclidean) IIB supergravity on the D$(-1)$ instanton background. We construct its holographic bulk realization as a one-dimensional maximal supergravity with 32 supercharges and local ${\rm SO}(10)$ invariance, capturing the full non-linear dynamics. Analyzing the bulk Killing spinor equations, we construct a general class of half-supersymmetric solutions, which typically break ${\rm SO}(10)$. We present their uplifts to IIB supergravity, and furthermore to pp-waves in twelve dimensions. 
These results provide the minimal setup for conducting precision tests of holography involving Einstein gravity. 
\end{abstract}

\maketitle

A particularly interesting corner of the holographic dualities is the correspondence between the backgrounds of D$p$-branes and their non-conformal field theory duals. Unlike the cases of conformal field theory duals, which are captured by the celebrated AdS/CFT correspondence, non-conformal dualities open a window into a less symmetric dynamics characterized by scale-dependent behavior. These dualities preserve sixteen supercharges and have been studied since the early days of holography \cite{Itzhaki:1998dd,Boonstra:1998mp}. In this context, the dual field theory is the reduction of ten-dimensional super Yang-Mills theory down to $p+1$ dimensions. 

Extrapolating this correspondence to the extremal case of $p=-1$ suggests a holographic duality between strings on the D$(-1)$ instanton background and super Yang-Mills theory reduced to zero dimensions. The latter is known as the IKKT matrix model, and has by itself been conjectured as a candidate for a nonperturbative formulation of type IIB superstring theory \cite{Ishibashi:1996xs}. The dual D$(-1)$ instanton background is a half-supersymmetric solution of Euclidean IIB supergravity with flat spacetime metric and non-trivial dilaton/axion  \cite{Gibbons:1995vg,Gubser:1996wt,Bergshoeff:1998ry,Ooguri:1998pf}. Despite the compelling and highly non-trivial results that have been obtained in the IKKT model \cite{Green:1997tn,Krauth:1998xh,Moore:1998et,Ambjorn:2000dx}, this holographic duality has remained largely unstudied, not at least due to various subtleties and degeneracies on both sides of the correspondence. However, recent work has started to probe holography in the polarized IKKT matrix model \cite{Hartnoll:2024csr,Komatsu:2024bop,Komatsu:2024ydh}, a supersymmetric ${\rm SO}(3) \times {\rm SO}(7)$ invariant mass deformation of the original model \cite{Bonelli:2002mb}.

In this letter, we explore the correspondence between gauge invariant operators of the IKKT model and Kaluza-Klein fluctuations around the D$(-1)$ instanton background. In particular, we consider the lowest BPS multiplet of such operators and identify its states with the lowest Kaluza-Klein fluctuations of IIB supergravity. We argue that their full non-linear dynamics is described by a one-dimensional maximal supergravity with 32 supercharges and local ${\rm SO}(10)$ invariance, which we construct explicitly. For a subsector of the theory, we present the explicit non-linear embedding into the Euclidean IIB theory. This paves the way for the holographic computation of correlation functions along the lines of  \cite{Kanitscheider:2008kd}.

We also study the Killing spinor equations of the bulk theory and determine a general family of half-supersymmetric solutions in terms of ten real constants. We provide their uplift to the IIB theory and recover the D$(-1)$ instanton as a special case. All these solutions turn out to have a flat spacetime metric in ten dimensions and can be further uplifted to twelve-dimensional pp-waves. We discuss their significance for the holographic description of supersymmetric deformations of the IKKT model.

Let us start from the Euclidean IKKT matrix model, obtained as the reduction of  
ten-dimensional super Yang-Mills theory to zero dimensions, with the action given by
\begin{equation}
S_{\text{\tiny{IKKT}}}=
-
\text{Tr}\Big[\tfrac14[X_a,X_b][X^a,X^b]-\tfrac12\bar{\Psi}\,\Gamma^a\,[X_a,\Psi]\Big]\,.\label{eq:IKKTaction}
\end{equation}
Here, $X_a$ and $\Psi^{\alpha}$ are bosonic and fermionic ${\rm SU}(N)$ matrices, respectively, transforming in the ${\bf 10}$ and the ${\bf 16}$ of ${\rm SO}(10)$, and $\Gamma^a$ denotes the $\mathrm{SO}(10)$ gamma matrices. To maintain clarity in the equations, we suppress spinor indices and we use $\bar \Psi_\alpha:=\Psi^\beta\mathcal C_{\alpha\beta}$ to represent the charge-conjugated spinor, where $\mathcal C$ is the symmetric charge conjugation matrix \footnote{Our charge conjugation matrix is related to that of \cite{Komatsu:2024ydh} by $\mathcal C_{\text{there}}=\mathrm{i}\,\mathcal C_{\text{here}}\,\Gamma_*$.}. Throughout the letter, we will use spinors with 32 components. However, in the IKKT model, the spinor $\Psi^\alpha$ and the supersymmetry parameter satisfy the chirality condition $\Gamma_* \Psi=\Psi$. The action (\ref{eq:IKKTaction}) is complex, but this is a standard feature of spinorial actions in Euclidean signature \cite{Krauth:1998xh}.

The IKKT action is invariant under ${\rm SU}(N)$, ${\rm SO}(10)$, and supersymmetry transformations \footnote{Although all these transformations are global symmetries in zero dimensions, we still refer to ${\rm SU}(N)$ as `gauge symmetries' in accordance with their higher-dimensional origin.}. The latter act as
\begin{equation}
\delta_\epsilon X^a=\bar\epsilon \,\Gamma^a\Psi\,,
\;\;\;\;
\delta_\epsilon \Psi=\frac12\Gamma^{ab}\epsilon\,[X_a,X_b]\,.
\end{equation}
The supersymmetry algebra closes on-shell into
\begin{equation}
[\delta_{\epsilon_1},\delta_{\epsilon_2}]
=\delta^{{\rm SU}(N)}_{\lambda}\,,\label{eq:IKKETalgebra}
\end{equation}
with the ${\rm SU}(N)$ parameter $\lambda=2\,\bar\epsilon_2\,\Gamma^b\,\epsilon_1X_b$\,.
As a consequence, supersymmetry transformations commute on the set of ${\rm SU}(N)$ invariant single-trace operators
\begin{equation}
{\cal O} = {\rm Tr}\left[ XX\dots\Psi\dots X \dots \Psi \dots \right]
\,.
\end{equation}
The set of all such operators, after dividing out the field equations of (\ref{eq:IKKTaction}),
organizes into supermultiplets \cite{Morales:2004xc}. These are all of the long type
\begin{equation}
{\cal R} \otimes 
\left(
{\bf 1}\oplus {\bf 16} \oplus\, {\bf 16} \wedge {\bf 16} \,\oplus\, {\bf 16} \wedge {\bf 16} \wedge {\bf 16} \,\oplus\, \dots
\right)
\;,
\end{equation}
with some ${\rm SO}(10)$ representation ${\cal R}$, except for a tower of protected BPS multiplets
\begin{equation}
\sum_{n=2}^\infty {\cal B}_n
\,.
\end{equation}
The latter are built from a lowest weight state
\begin{equation}
    {\cal O}^{a_1 \dots a_n} =
    {\rm Tr}\big[ X^{(\!(a_1}X^{a_2} \dots X^{a_n)\!)} \big]
    \,,
\end{equation}
and combine the ${\rm SO}(10)$ representations
\begin{align}
{\cal B}_n =\,& 
[n,0000]_n
\oplus [n\!-\!1,0001]_{n+\frac12}
\oplus [n\!-\!2,0100]_{n+1}
\oplus 
\nonumber\\
&{}
[n\!-\!3,1010]_{n+\frac32}
\oplus [n\!-\!3,0020]_{n+2}
\oplus [n\!-\!4,2000]_{n+2}
\nonumber\\
&{}
\oplus [n\!-\!4,1010]_{n+\frac52}
\oplus [n\!-\!4,0100]_{n+3}
\nonumber\\
&{}
\oplus [n\!-\!4,0001]_{n+\frac72}
\oplus [n\!-\!4,0000]_{n+4}
\,,
\label{eq:BPS}
\end{align}
with the subscript counting the
naive dimension $\Delta$, corresponding to $\Delta_X=1, \Delta_\Psi=\frac32$\,. 
For generic $n$, the BPS multiplet ${\cal B}_n$ carries
 \begin{equation}
\#_{\rm bos}  = 1+128\,{\rm dim}\,{\cal R}_{[n-2,0,0,0]} = 1+\#_{\rm ferm}\,,
\label{eq:dof}
\end{equation}
degrees of freedom. 
For the lowest BPS multiplet ${\cal B}_2$, the generic structure (\ref{eq:BPS}) 
degenerates into
\begin{align}
{\cal B}_2 =\,& 
{\bf 54}_{+2} \oplus
{\bf 144}_{+\frac52} \oplus
{\bf 120}_{+3}
\ominus
{\bf 45}_{+4} 
\ominus
{\bf 16}_{+\frac92}
 \,,
 \label{eq:BPS2}
\end{align}
consistent with the counting (\ref{eq:dof}).
While in higher dimensions states of negative multiplicity in the lowest multiplet correspond to the Goldstone modes to be subtracted from a massive state, here they account for the global ${\rm SO}(10)$ and supersymmetry, respectively.
The first three terms in (\ref{eq:BPS2}) correspond to the operators
\begin{align}
{\cal O}^{ab} =\,& {\rm Tr}[X^{a} X^{b}]-\tfrac1{10}\,\delta^{ab}\,{\rm Tr}[X^{c} X^{c}] 
\,,\label{eq:O1}\\
{\cal O}^a =\,& {\rm Tr}[X^a\,\Psi] - \tfrac19\, {\rm Tr}[X_b\,\Gamma^{ab} \Psi]
\,,\label{eq:O2}\\
{\cal O}^{abc} =\,& {\rm Tr}\left[X^{a} [X^{b},X^c]\right] -\tfrac18\, {\rm Tr}\left[\bar\Psi \Gamma^{abc} \Psi \right]
\,,\label{eq:O3}
\end{align}
on which the supersymmetry algebra (\ref{eq:IKKETalgebra}) closes with
\begin{align}
\delta_\epsilon \mathcal O^{ab}=\,&\tfrac95\, \bar \epsilon\,\Gamma^{(a}\mathcal O^{b)}\,,\\
    \delta_\epsilon \mathcal O^a=\,&\tfrac{1}{18}\left(
    7\,\Gamma_{bc}\,\epsilon \,\mathcal O^{abc}-\Gamma^{abcd}\,\epsilon\,\mathcal O_{bcd}\right)
    \,,
\end{align}
and $\delta_\epsilon {\cal O}^{abc}=0$\,. As a consequence of the nilpotent structure of the supersymmetry transformations, the vev deformations of the IKKT model (\ref{eq:IKKTaction}) by components of ${\cal O}^{ab}$ preserve all 16 supercharges. 

In the higher-dimensional holographic dualities, it can be shown that the BPS part of the SYM spectrum in $p+1$ dimensions precisely coincides with the full spectrum of supergravity fluctuations around the D$p$-brane near horizon background \cite{Morales:2004xc}. In particular, in the known higher-dimensional cases, the lowest BPS multiplet in the Yang-Mills spectrum is realized as the lowest Kaluza-Klein supergravity multiplet in the bulk. In turn, its dynamics is described by a maximal gauged supergravity in $(p+2)$ dimensions, with gauge group ${\rm SO}(9-p)$, obtained by reduction from ten dimensions on $S^{8-p}$ \cite{Boonstra:1998mp}.
%In the higher-dimensional holographic dualities, the lowest BPS multiplet in the Yang-Mills spectrum is realized as the lowest Kaluza-Klein supergravity multiplet in the bulk. 
%
Accordingly, for IKKT, the associated bulk theory should correspond to a one-dimensional ${\rm SO}(10)$-gauged supergravity combining the lowest Kaluza-Klein fluctuations (\ref{eq:BPS2}) around the D$(-1)$-instanton background. In the following, we construct this model explicitly.

We begin by presenting the dual half-supersymmetric instanton solution of Euclidean IIB supergravity \cite{Gibbons:1995vg,Gubser:1996wt}. The metric in the Einstein frame, the axion $\chi$, and the dilaton $\Phi$ are given by
\begin{equation}
d s^2_{10}=dr^2+r^2\,d\Omega_9^2\,,\qquad
e^{\Phi}=H=-\chi^{-1} \,,
\label{eq:D-1brane}
\end{equation}
while all the other p-forms vanish. The harmonic function $H$ reads
\begin{equation}
H=h+\frac{Q}{r^8}\;.
\end{equation}
Here, $d\Omega_9$ is the line element on the unit 9-sphere, such that the metric is flat and that the solution has an $\rm{SO}(10)$ isometry. The constant $Q$ is proportional to the instanton charge, while the constant $h$ can be shifted by the ${\rm SL}(2,\mathbb{R})$ symmetry of the IIB theory \cite{Bergshoeff:1998ry}. Note that the sphere radius $r$, which plays the role of Euclidean time, serves as the holographic coordinate. The string frame metric describes a wormhole geometry, but becomes flat in the near-horizon limit \cite{Ooguri:1998pf}. This contrasts with D$p$-brane near-horizon geometries for $p\geq 0$ which correspond to domain wall solutions \cite{Boonstra:1998mp}.

To construct the maximally supersymmetric theory that governs the lowest Kaluza-Klein fluctuations around this background, we will first construct a (non-supersymmetric) consistent truncation of IIB Euclidean supergravity on the sphere $S^9$ as a special case of a more general consistent $S^d$ truncation of $(d+1)$-dimensional dilaton-axion gravity. We then extend the resulting one-dimensional theory for $d=9$ to include the full field content of the supermultiplet (\ref{eq:BPS2}) by imposing maximal supersymmetry.

The starting point of our construction is the $(d+1)$-dimensional bosonic Lagrangian 
\begin{equation}
{\cal L}_{d+1} =
 |E|\Big(R - \tfrac12 \partial_\mu \Phi \,\partial^\mu\Phi  
+\tfrac12 e^{2\Phi}\, \partial_\mu \chi \,\partial^\mu \chi \Big) \;,\label{eq:LEu}
\end{equation}
in Euclidean spacetime. We keep the dimension arbitrary for now, but later focus on the $d=9$ case, where the Lagrangian \eqref{eq:LEu} describes the axion/dilaton sector of Euclidean IIB supergravity \footnote{Up to a boundary term, which does not affect our analysis that only relies on the field equations}. Note that all fields in this action are real \footnote{In principle, the fields in the Euclidean IIB theory and in the one-dimensional supergravity \eqref{eq:SUGRAL} are complex, as SO(10) Majorana spinors do not exist. However, the supersymmetric Lagrangians are holomorphic functions of these fields, and the supersymmetry variations act holomorphically. As argued in \cite{Nicolai:1978vc, Krauth:1998xh, Hartnoll:2024csr}, this structure allows one to choose a real section and consistently treat all fields as real.}, but in Euclidean signature the sign of the axion kinetic term has changed \cite{Gibbons:1995vg}. Consequently, the coset parametrized by $\Phi$ and $\chi$ is $\rm{SL}(2)/\rm{SO}(1,1)$. This relative sign is essential for ensuring that the energy momentum tensor of the flat space solution \eqref{eq:D-1brane} vanishes.

When expanded around the sphere $S^{d}$, the $(d+1)$-dimensional theory \eqref{eq:LEu} admits a consistent truncation to a finite set of Kaluza-Klein modes. The non-linear dynamics of these modes are governed by a one-dimensional theory, which is outlined below. The Kaluza-Klein truncation Ansatz for the $(d+1)$-dimensional fields is inspired by the Ans\"{a}tze of \cite{Cvetic:2000dm,Ciceri:2023bul} and reads
\begin{align}
d s_{D}^2 =\,&  e^{\tfrac{8d\phi}{d-1}}\,\Delta\, \mathrm{e}^2\,dt^2 
+ g^{-2}\, e^{\tfrac{8\phi}{d-1}} \, {T}^{-1}_{ij}\, 
{ D}\mu^i\, { D}\mu^j
\,,\label{eq:Ansatz1}\\[3mm]
e^ {\Phi} =\,&  
e^{-4\phi}\,\Delta^{-1}
\,,\label{eq:Ansatz2}\\
\chi=&-\frac1{2\,\mathrm{e}g}
\left(
T^{-1}_{ij}\, D_t T_{kj}\,\mu^{i}\mu^{k}+\frac{8\,\dot\phi}{1-d}\right)\,,\label{eq:Ansatz3}
\end{align}
where the embedding coordinates of the unit $S^d$-sphere $\mu^i$, with $i=1,\ldots,d+1$, satisfy $\mu^i\mu^i$=1, and where $\Delta=T_{ij}\,\mu^i\mu^j$, The symmetric, positive definite matrix $T_{ij}\in \mathrm{SL}(d+1)$ contains the lower-dimensional scalar fields that only depend on the Euclidean time coordinate $t$. The other one-dimensional fields include the Einbein $\rm e$, a dilaton $\phi$, and a set of $\mathrm{SO}(d+1)$ gauge fields $A_{[ij]}$. The latter appear implicitly through the gauge covariant differential $  D$, which acts as follows
\begin{align}
{ D} \mu^i =\,& d\mu^i - g\, A_{ji}\,\mu^j\,dt\,,\\
{ D}_t  T_{ij} =\,&
  \dot{T}_{ij} -2\, g\,A_{k(i}  T_{j)k}\,,
\label{eq:covD}
\end{align}
with $ D T_{ij}=: D_t T_{ij}\,dt$ and where $g$ is the gauge coupling constant. The dot represents the derivative with respect to the time $t$. The local $\mathrm{SO}(d+1)$ invariance of the resulting one-dimensional gravity theory is inherited from the isometry group of the sphere $S^d$. Its Lagrangian reads 
\begin{align}
\mathcal L_1 =\,& \frac{8(d+1)}{(d-1)}\mathrm{e}^{-1}\, \dot \phi^2+\frac14 \mathrm{e}^{-1}\, D_t{T}_{ij}^{-1}\,  D_t\,{T}_{ij}
\nonumber\\
&
-\frac12\,\mathrm{e}\, g^2 \,e^{8\phi}\left(
2\,T_{ij}T_{ij}-( T_{ii})^2\right)\,.
\label{eq:L1Eu}
\end{align}
The scalar potential appearing in the second line is characteristic of sphere truncations of gravity theories. To explicitly prove the consistency of this Kaluza-Klein truncation, it is necessary to show that, after substitution of the truncation Ansatz, the $(d+1)$-dimensional field equations reduce to those derived from the one-dimensional Lagrangian \eqref{eq:L1Eu}. We have checked that this is indeed the case and the details of the computation will be presented elsewhere \cite{Ciceri:2025wpb}.

The field equations deriving from \eqref{eq:L1Eu} can be conveniently expressed in an $\mathrm{SO}(d+1)$ gauge where $T_{ij}=\delta_{ij}\,e^{\varphi_j}$, with $\sum_{i}\varphi_i=0$. Additionally, we exploit the time reparametrization invariance of the theory to set $\mathrm{e}=1$. Varying with respect to the Einbein, the dilaton, and the scalars $\varphi_i$, respectively, leads to 
\begin{align}
\frac{16(d+1)}{d-1}\,\dot\phi^2
  =\,&\frac12\,\sum_k \dot\varphi_k^2
  -g^2\,e^{8\phi}\,V_0\,,\label{eq:EOME}\\
\frac{4(d+1)}{d-1}\,\ddot\phi
=\,&-g^2\,e^{8\phi}\,V_0\,,\label{eq:EOMphi}\\[1mm]
 \frac12\,\ddot\varphi_i -\frac{4}{d-1}\,\ddot\phi
  =\,&
g^2\,e^{8\phi}\,
\Big(
2\,e^{2\varphi_i}- \sum_k e^{\varphi_i+\varphi_k}
\Big)
 \,,
 \label{eq:EOMvphi}
 \end{align}
with the potential
\begin{equation}
V_0 = 
2\,\sum_k e^{2\varphi_k} - \Big(\sum_k e^{\varphi_k}\Big)^2 
\,.
\end{equation}
The gauge fields $A_{ij}$ are set to zero by their own field equations, or decouple from the dynamics. For vanishing $\varphi_i$, the system admits an $\mathrm{SO}(d+1)$-invariant solution, which for $d=9$ uplifts to the D-instanton \eqref{eq:D-1brane} via the truncation Ansatz (\ref{eq:Ansatz1})--(\ref{eq:Ansatz3}). This solution actually belongs to a broader class of half-supersymmetric solutions, which we will derive below.

As discussed above, we expect the holographic duality to relate the IKKT operators \eqref{eq:O1}--\eqref{eq:O3} to the lowest Kaluza-Klein fluctuations of the full IIB supergravity, corresponding to the multiplet (\ref{eq:BPS2}). In principle, one should thus extend the $S^9$ truncation Ansatz \eqref{eq:Ansatz1}--\eqref{eq:Ansatz3} to include all the ten-dimensional p-forms. This would likely require elaborate techniques based on exceptional geometry that are beyond the scope of this letter. We instead opt to directly construct the maximally supersymmetric extension of the one-dimensional Lagrangian \eqref{eq:L1Eu} for $d=9$. 

Due to the triviality of the tangent space in one dimension, the fermionic fields of the resulting  supergravity theory simply consist of Grassman variables transforming in different $\mathrm{SO}(10)$ representations. They include the gravitino $\psi^\alpha$, as well as the fermions $\lambda^\alpha$ and $\chi^\alpha_a$. We employ here the $\mathrm{SO}(10)$ spinor notations introduced in \eqref{eq:IKKTaction}. Note, however, that unlike the chiral spinor $\Psi$ in the IKKT model, a supergravity spinor such as the gravitino $\psi^\alpha$ (or the supersymmetry parameter $\epsilon^\alpha$) carries 32 independent components. Likewise, the vector-spinor $\chi^\alpha_a$ satisfies the trace condition $\Gamma^a\,\chi_a=0$ and carries $2\times 144$ components. This indeed is the proper off-shell bulk content to realize the on-shell multiplet (\ref{eq:BPS2}). %As for the IKKT model, the one-dimensional action is complex \textcolor{red}{due to the SO(10) gamma matrices appearing in the fermionic couplings}.

In addition to the fermionic fields, maximal supersymmetry also necessitates enlarging the bosonic sector by adding 120 axion fields $a_{[ijk]}$, according to (\ref{eq:BPS2}). Up to second order in the axions and in the fermion fields, the resulting supergravity Lagrangian takes the form \footnote{To be precise, to this order in the fermions, maximal supersymmetry admits a second extension of \eqref{eq:L1Eu} with in particular an opposite sign in the kinetic and the potential term for the axion fields $a_{ijk}$, and additional $\Gamma_*$ matrices in the axion-fermion couplings.}
\begingroup
\allowdisplaybreaks
\begin{widetext}
\begin{align}
\mathcal L_1 =\,& 10\,\mathrm{e}^{-1}\, \dot \phi^2+\frac14 \mathrm{e}^{-1}\, D_t\,{T}_{ij}^{-1}\,  D_t\,{T}_{ij}-\frac1{12}\mathrm{e}^{-1}\,e^{-2\phi}\, T^{-1}_{ij}T^{-1}_{kl}T^{-1}_{mn}\, D_t a_{ikm}\, D_t a_{jln}
+20\,\bar\lambda\,\mathcal D_t\lambda+2\,\bar\chi^a\, \mathcal D_t\chi_a
\nonumber\\[1ex]
\,&-\frac12\,\mathrm{e}\, g^2 \,e^{8\phi}\left(
2\,T_{ij}T_{ij}-( T_{ii})^2+\frac12 e^{-2\phi}
\left(a_{ijk}\,a_{lmn}\,{T}_{il}{ T}^{-1}_{jm}T^{-1}_{kn}-2\,a_{ijk}\,a_{ijl}\,{T}^{-1}_{kl}\right)\right)
\nonumber\\[1ex]
\,&
+4\,g\,\mathcal X \,\bar\psi\, \lambda
-2\,g\,\mathcal X^{ab} \,\bar\psi\, \Gamma_b\Gamma_*\,\chi_a-21\,\mathrm{e}\,g\,\mathcal X \,\bar\lambda\, \Gamma_*\,\lambda
-16\,\mathrm{e}\,g\,\mathcal X^{ab}\,\bar\lambda\, \Gamma_b\,\chi_a
-\mathrm{e}\,g \left( 4\,\mathcal X^{ab}-\frac{1}{10}\delta^{ab}\,\mathcal X \right)\,
\bar\chi_a\,\Gamma_*\,\chi_b
\nonumber\\[1ex]
\,&
-20\,\mathrm{e}^{-1}\bar\psi\,\Gamma_*\lambda\,\dot\phi
+2\,\mathrm{e}^{-1}\bar\psi\,\Gamma^{b}\chi^{a}\,\mathcal P_{ab}-\frac{1}{2}\mathrm{e}^{-1}\,e^{-\phi}\,\bar\chi^{a}\,\Gamma^{bc}\,\psi\,p_{abc}
-\frac{1}{12}\,e^{-\phi}\,\,\bar\chi^{a}\,\Gamma^{bcd}\,\chi^a\,p_{bcd}-e^{-\phi}\,\bar\chi^a\,\Gamma^b\,\chi^c\,p_{abc}
\nonumber\\[1ex]
\,&{}
-\frac{1}{6}\,\mathrm{e}^{-1}\,e^{-\phi}\,\bar\lambda\,\Gamma^{abc}\,\Gamma_*\psi\,p_{abc}-e^{-\phi}\,\bar\lambda\,\Gamma^{abc}\lambda\,p_{abc}-e^{-\phi}\,\bar \chi^c\,\Gamma^{ab}\Gamma_*\,\lambda\,p_{abc}
+\mathcal L_{\rm Yuk}[a_{ijk}]
\,,
\label{eq:SUGRAL}
\end{align}
\end{widetext}
\endgroup
\noindent where $D_t a_{ijk}=\dot a_{ijk}-3\,g\,A_{l[i}\,a_{jk]l}$.  The first two terms of the first two lines coincide with the non-supersymmetric Lagrangian \eqref{eq:L1Eu}, while the Yukawa couplings in the third line involve the scalar combinations 
\begin{equation}
\mathcal X^{ab}:=e^{4\phi}\Big(\mathcal T^{ab}-\frac{1}{10}\delta^{ab}\,\mathcal T^{cc}\Big)\,,\;\;\;\mathcal X:=e^{4\phi} \,\mathcal T^{aa}\,,
\end{equation}
that transform in the $\bf{54}$ and $\bf 1$ of $\mathrm{SO}(10)$, respectively. Here, we have introduced the $\mathrm{SO}(10)$ counterpart $\mathcal{T}^{ab}$ of the scalar matrix $T_{ij}$. These unimodular symmetric matrices are both expressed as
\begin{equation}
T_{ij}=V_{i}{}^a\,V_{j}{}^a\,,\;\;\;\;\;\mathcal T^{ab}=V_i{}^a\,V_i{}^b\,,
\end{equation}  
in terms of a representative $V_i{}^a\in \mathrm{SL}(10)$ for the coset space $\mathrm{SL}(10)/\mathrm{SO}(10)$. The associated (gauge covariant) Maurer-Cartan current decomposes into 
\begin{align}
(V^{-1})_a{}^i D_t V_{i\,b}&=\,(V^{-1})_a{}^i \,(\dot V_{i\,b}-g\,A_{ji}V_{j\,b})\nonumber\\
&=:\,\mathcal P_{(ab)}+\mathcal{Q}_{[ab]}\,,
\end{align}
where the component $\mathcal P_{ab}$ naturally couples to the fermions in the Lagrangian and in the supersymmetry transformations. The composite connection $\mathcal Q_{ab}$ enters the definition of the $\mathrm{SO}(10)$ covariant derivative $\mathcal D_t$ that acts on fermions. Likewise, the axions couple to the fermions via the dressed current
\begin{equation}
    p_{abc}:=
     (V^{-1})_{a}{}^{i}(V^{-1})_{b}{}^{j}(V^{-1})_{c}{}^{k}\,
    D_t a_{ijk}\,.
\end{equation}
Note that the absence of a kinetic term for the gravitino in the Lagrangian is expected, as the standard Rarita-Schwinger term vanishes in one dimension. Note also that in the last line of \eqref{eq:SUGRAL}, to keep the equations manageable, we have refrained from presenting explicitly the Yukawa couplings involving the axion fields. Comprehensive results, including higher-order terms in the axions, will be provided in a follow up publication \cite{Ciceri:2025wpb}.

The supergravity Lagrangian \eqref{eq:SUGRAL} is invariant under local $\mathrm{SO}(10)$ transformations and supersymmetry. For the aims of this letter, it is sufficient to present the supersymmetry variations of the fields in the sector where $a_{ijk}=0$. The bosons in this truncation transform as 
\begin{align}
    \delta_\epsilon \mathrm e=\,&\bar \epsilon \,\psi\,,\qquad
    \delta_\epsilon \phi=\, \bar \epsilon\, \Gamma_* \,\lambda\,,\\[0.5ex]
    \delta_\epsilon V_i{}^a=\,&\frac12 \,\bar\epsilon \,\Gamma^{(a}\,\chi^{b)}\,V_i{}^b\,,\\[0.5ex]
    \delta_\epsilon A_{ij}=\,&
e^{4\phi}\,\mathrm{e}\,\Big(4\,\bar\epsilon\,\Gamma^{ab}\,\lambda
-2\,\bar\epsilon\,\Gamma^{a}\Gamma_*\,\chi^b\Big)\,
V_{ij}{}^{ab}
\nonumber\\[0.5ex]
&{} +e^{4\phi}\,
\bar\epsilon\,\Gamma^{ab}\Gamma_*\,\psi\,V_{ij}{}^{ab}
\,,
\end{align}
with $V_{ij}{}^{ab}:=V_{[i}{}^aV_{j]}{}^b$, while the fermions transformations read
\begin{align}
\delta_\epsilon \psi=\,& \mathcal D_t\,\epsilon-\frac14\,g\,\mathrm{e}\,\mathcal X\,\Gamma_*\,\epsilon\,,\\[0ex]
\delta_\epsilon\lambda=\,&\frac{1}{2}\mathrm{e}^{-1}\dot\phi\,\Gamma_{*}\,\epsilon+\frac{1}{10}\,g\,\mathcal X\,\epsilon\,,\\[0ex]
\delta_\epsilon\chi_{a}=\,&\frac{1}{2}\mathrm{e}^{-1}\Gamma^{b}\,\epsilon\,\mathcal P_{ab}-\frac12\,g\,\mathcal X_{ab}\,\Gamma^b\Gamma_*\epsilon\,.
\end{align}
In this sector, it becomes straightforward to study the Killing spinor equations. The gamma matrix structure implies that supersymmetry can either be broken completely, or by half, where in the latter case $\epsilon^\alpha$ is restricted by $\Gamma_*\, \epsilon = \pm\epsilon$. We focus on the positive chirality solutions to align with the chirality convention of the IKKT model spinor. In the diffeomorphisms and $\mathrm{SO}(10)$ gauges adopted previously, where $\mathrm{e=1}$ and $V_i{}^a=\delta_i^a \,e^{\varphi_i/2}$, and upon redefining
\begin{equation}
    \phi_i := \varphi_i -\phi\,,\;\;\;\;\;\;\;e^\phi=\prod_i e^{-\phi_i/10} \,,
\end{equation}
the first order Killing equations reduce to
\begin{equation}
    \partial_{\tau} {\phi}_i    = -e^{\phi_i}
    \quad\Longrightarrow\quad
    e^{\phi_i}=\frac{1}{\tau+c_i}\,,
    \label{eq:phisol}
\end{equation}
with real constants $c_i$. It is straightforward to verify that this yields a solution to the equations of motion \eqref{eq:EOME}--\eqref{eq:EOMvphi} for any choice of the $c_i$. Here, $\tau(t)$ is a redefined time coordinate that satisfies $\dot \tau=-2\,g\,e^{5\phi}$.

The uplifts of these solutions to half-supersymmetric Euclidean IIB supergravity solutions are obtained using the truncation Ansatz \eqref{eq:Ansatz1}--\eqref{eq:Ansatz3}, and read
\begin{align}
ds_{10}^2 =\,&  \frac{1}{4\,g^{2}}\,\left(\mu^i\mu^i e^{\phi_i}\right)\, d\tau^2 
+ \,d\mu^i d\mu^i\,\frac{e^{-\phi_i}}{g^2}
\,,\label{eq:upliftMetric}\\
e^{\Phi} =\,&
e^{-5\phi}\left(\mu^i\mu^i\,e^{\phi_i}\right)^{-1} = -\chi^{-1}
\,.
\label{eq:upliftSusy}
\end{align}
with the explicit form of the functions $\phi_i$ in \eqref{eq:phisol}, and where all the other IIB p-forms vanish. For any values of the constants $c_i$, the IIB metric (\ref{eq:upliftMetric}) still turns out to be flat in the Einstein frame. When all the constants $c_i$ are equal, the solution reduces to the $\mathrm{SO}(10)$ invariant D($-1$) background \eqref{eq:D-1brane} (in the near-horizon limit \cite{Ooguri:1998pf}) that is dual to the IKKT model. More generally, for $I$ sets of $n_I\geq 2$ equal constants, we expect these solutions to be the analogues of the smeared D3-brane solutions studied in \cite{Freedman:1999gk}, describing $\prod_I {\rm SO}(n_I)$-invariant distributions of $N$ instantons spread across the nine transverse directions. They should be dual to the supersymmetric vev deformations of the IKKT model, associated with the corresponding components of the operator \eqref{eq:O1}. Finally, following \cite{Tseytlin:1996ne, Tseytlin:1997ps}, the flat solutions \eqref{eq:upliftSusy} can be further uplifted on a torus to the following pp-wave solutions of pure Lorentzian gravity in twelve-dimensions,
\begin{equation}
    ds_{12}^2=dx_+dx_- +e^\Phi dx_- dx_-+ds_{10}^2\,,
\end{equation}
with the light-cone coordinates $x_\pm=x_{12}\pm x_{11}$, and where $x_{11}$ is the Lorentzian time.

In this letter, we have constructed maximal supergravity in one dimension, finally completing the list of maximal supergravities across all spacetime dimensions. The model, by itself, stands out as a new and distinguished example of supersymmetric quantum mechanics, featuring 32 supercharges, and an intricate highly non-linear scalar potential which deserves further exploration. In particular, its quantization could be approached using techniques that bypass the technical difficulties of quantum field theory. It would also be interesting to explore whether the non-perturbative regime of the model can be studied using supersymmetric localization.

As a maximal gauged supergravity, this model is distinct in that it specifically builds on a sphere compactification without an obvious higher-dimensional origin for the limit $g\rightarrow0$ of vanishing coupling constant. In maximal supergravities in higher dimensions $D>1$, the limit $g\rightarrow0$ reproduces the maximal supergravity obtained by toroidal reduction with a symmetry enhancement to the rank $(11-D)$ exceptional group ${\rm E}_{11-D}$. For the present model this would suggest a highly non-trivial realization of the hyperbolic exceptional group ${\rm E}_{10}$ on the $g\rightarrow0$ limit of (\ref{eq:SUGRAL}), in line with the longstanding conjecture of \cite{Julia:1982gx}. One may expect further intriguing connections to the constructions of \cite{Mizoguchi:1997si,Damour:2002cu}.

From the holographic perspective, the model yields a gravitational (bulk) realization of the lowest BPS multiplet of gauge invariant operators in the IKKT matrix model. As such, it sets the stage for a holographic computation of correlation functions among the operators \eqref{eq:O1}--\eqref{eq:O3} along the lines of \cite{Kanitscheider:2008kd}. This offers a new angle on the large $N$ limit of IKKT which has many subtleties on its own. It would be most interesting to compare this to the existing results in this model 
\cite{Green:1997tn,Krauth:1998xh,Moore:1998et,Ambjorn:2000dx}. Another ambitious question is the existence of a Lorentzian analogue of our gravitational dual whose solutions seem to require Euclidean signature. In particular, this would have to capture the qualitative differences of the Lorentzian matrix model, revealed in recent numerical simulations \cite{Asano:2024def,Chou:2025moy}.

We have also in this letter constructed a general class of 1/2-BPS solutions of the model, which all uplift to IIB solutions with flat spacetime metric, and preserve all the supersymmetries of the D$(-1)$ background. We expect them to be dual to states in the `Coulomb branch' of the IKKT model, and it would be interesting to study the corresponding distributions of instantons in more detail. It will also be interesting to explore potential relations to known operators deformations of the IKKT model, in particular to the so-called polarized IKKT matrix model \cite{Bonelli:2002mb} for which recent progress has been achieved in \cite{Hartnoll:2024csr,Komatsu:2024ydh,Komatsu:2024bop}, as well as to the deformations that are dual to the spherical branes solutions of \cite{Bobev:2018ugk,Bobev:2024gqg}. Another recently explored deformation is based on holography for the D($-1$)/D7 system \cite{Billo:2021xzh} whose near-horizon background shares many similarities with our solutions \cite{Aguilar-Gutierrez:2022kvk}.

Let us finally note, that the holographic duality discussed here is rather special in that the IIB spacetime metric of the D$(-1)$ instanton background is actually flat. Yet, the identification of states dual to the matrix model operators builds on the expansion of supergravity fields into $S^9$ sphere harmonics, just as in higher dimensions, and close to the standard AdS/CFT correspondence. It is tempting to speculate whether this model may in fact hold lessons for flat space holography.

\medskip
\begin{acknowledgments}
We would like to thank Guillaume Bossard, Simeon Hellerman, Axel Kleinschmidt, and Ergin Sezgin for useful discussions. We also thank Nikolay Bobev, Pieter Bomans and Fridrik Freyr Gautason for useful comments on the first version of this letter. We also thank Adrien Martina for pointing out a typo in the uplift formulae. The research of FC is funded by the Deutsche Forschungsgemeinschaft (DFG, German Research Foundation) – Project number: 521509185.
\end{acknowledgments}


%\bibliographystyle{apsrev4-2}
%\bibliography{refs}

\end{document}